\newcount\style \style=2
\ifodd\style
\documentstyle[aas2pp4]{article}
\else
\documentstyle[12pt,aasms4]{article}
\fi 

\begin{document}

\setlength{\baselineskip}{12pt}

\newcommand\bb[1] {   \mbox{\boldmath{$#1$}}  }
\newcommand\del{\bb{\nabla}}
\newcommand\bcdot{\bb{\cdot}}
\newcommand\btimes{\bb{\times}}
\newcommand\vv{\bb{v}}
\newcommand\B{\bb{B}}
\newcommand\BV{Brunt-V\"ais\"al\"a\ }
\newcommand\iw{ i \omega }
\newcommand\kva{ \bb{k\cdot v_A}  }
\newcommand\beq{\begin{equation}}
\newcommand\eeq{\end{equation}}

\def\dd{\partial}
\lefthead{Fromang \& Balbus}
\righthead{When is Uniform Rotation an Energy Minimum?}
\slugcomment{Submitted to ApJ.}

\title{When is Uniform Rotation an Energy Minimum?}

\author{S\'ebastien Fromang}
\affil{Institut d'Astrophysique de Paris, 98 bis Boulevard
Arago, \\ 75014 Paris, France\  fromang@iap.fr}
\medskip

\author{and}

\author{ Steven A. Balbus}
\affil{Dept. of Astronomy, University of Virginia, P.O.~Box
3818,\\ Charlottesville VA 22903, USA\  sb@virginia.edu}

\begin{abstract}

A simple variational calculation is presented showing that a uniformly
rotating barotropic fluid in an external potential attains a true
energy minimum if and only if the rotation profile is everywhere
subsonic.  If regions of supersonic rotation are present, fluid
variations exist that could take the sytem to states of lower energy.
In any given system, these states may or may not be dynamically
accessible, but their existence is important.  It means that extending
the degrees of freedom available to the fluid (say by weak magnetic
fields) may open a path to fluid instabilities.  Whether astrophysical
gaseous nebula tend toward states of uniform rotation or toward more
Keplerian core-disk systems appears to be largely a matter of whether
the rotation profile is transonic or not.  The suggestion is made that
the length scale associated with coherent molecular cloud cores is
related to the requirement that the cores be stable and rotate
subsonically.

\end{abstract}


\section{Introduction}

It is a classical notion of hydrodynamics that a state of uniform
rotation represents an energy minimum.  A simple discussion of this
is given by Lynden-Bell \& Pringle (1974).  The very
same paper notes, however, that accretion disks behave in the opposite
sense of seeking a state of uniform rotation.  Instead, they strive to
segregate matter and angular momentum, hoarding all the mass at their
centers, and expelling all the angular momentum to large distances.
Lynden-Bell and Pringle showed that this behavior emerges as a
consequence of mass exchange in the disk, so that whatever unfavorable
energy cost is incurred by the differential rotation is more than
offset by the decrease in gravitational potential energy.

The question arises of when, more generally, a rotating system system
will tend toward a state of constant angular velocity, and when it will
accrete toward the singular ``black hole'' state.  Do systems with
mass exchange in a gravitational potential necessarily lead to a black
hole final state?  The problem is clearly one of widespread
astrophysical interest, as it has a direct bearing on when rotating gas
coalesces into cores and when it simply rotates uniformly.

The magnetorotational instability (Balbus \& Hawley 1998) is
a natural mechanism to drive a rotating system to either of these
states.  Imagine a rotating cloud with a small angular velocity
gradient.  The singular state will be attained only if the instability
acts continuously without turning itself off.  But this is precisely
what the outcome of a core collapse would be:  should the core contract
due to angular momentum loss, it would rotate more rapidly and increase
the initial angular velocity gradient.  This in turn increases the
growth rate of the instability, leading to an even greater outward flow
of angular momentum from the core.  On the other hand, should the
extracted angular momentum bring the more slowly rotating outer zones
into uniform rotation with the more rapidly rotating core, the
instability would cease.  What then are the conditions that lead a system
down one path or the other?

In this {\it Letter} we derive a very simple, general result of some
astrophysical significance.  Though our treatment is brief and
straightforward, we are unaware of any discussion of the result in the
prior astrophysical literature.  We consider a uniformly rotating
barotropic gas confined by an {\em external} potential.  The case of
{\em self-gravitating} cylinders has been investigated by Inagaki \&
Hachisu (1978) and Hachisu (1979), but is significantly more complex
than our treatment.  (A comparison of our work with the results of
these authors will be presented in a later more detailed numerical and
analytic study.)  We show that the energy is a true minimum in a state
of uniform rotation if and only if the rotation velocity is everywhere
less than the sound speed.  If there exists a region of supersonic
rotation, then it is possible to find second-order variations that will
lower the energy from the extremal state.  In this case, the uniformly
rotating state is not a minimum, but a saddle point.  Supersonic
rotation, when it is unstable, evolves not toward a uniform profile,
but towards a state of central mass concentration with a surrounding
Keplerian-like disk.

The plan of this {\em Letter} is as follows.  In \S 2, we present a
variational treatment of a non--self-gravitating, uniformly rotating
barotropic fluid as described above.  In \S 3, we present a brief
discussion of the astrophysical significance of our results, arguing
that they provide a simple way of deciding when systems evolve toward
uniform rotation, and when they evolve toward a strong central mass
with a surrounding disk--halo structure.  We also show that our
findings may be relevant to understanding the length scales over which
molecular cloud cores show coherence.

\section {Analysis}

\subsection{Equilibrium First Order Variation}

Consider a rotating barotropic gas confined by an external potential $\Phi$.
The energy of the system is
\beq
E = \int \left( {1\over 2} \rho v^2 +\varepsilon + \rho \Phi
\right) \, dV
\eeq
where $\rho$ is the mass density, $v$ the velocity, and $\varepsilon$
is the internal energy density.  Introduce first order Eulerian
variations into the fluid quantities (e.g. $\rho\rightarrow \rho
+{\delta\rho}$, etc.) subject to the constraint that the total mass and
the total angular momentum remain constant.  We use Lagrangian
multipliers $-{\zeta}$ and $-\Omega$ to ensure these
conditions (the minus
signs are for later convenience), and vary the quantity
\beq
I = E - {\zeta}\int \rho dV - \Omega \int \rho R v dV,
\eeq
where $R$ is the cylindrical radius about the axis of rotation.
If the energy is an extremum, then $\delta I = 0$, or
\beq
0 = \int \left[ \delta\rho\,\left(
{-\zeta}- \Omega R v+ {\delta ({\rm int}) \over \delta \rho}\right)
+ \delta v\, \rho \left(  v  - \Omega R \right)\right] dV, 
\eeq
where the variational derivative of the integrand of equation (1) is
\beq
{\delta ({\rm int}) \over \delta \rho} = 
{v^2\over2} + {d\varepsilon\over d\rho} + \Phi.
\eeq
We require the coefficients of $\delta v$ and
$\delta\rho$ to vanish independently.  The first
condition leads to solid body rotation,
\beq\label{fixomega}
v = \Omega R, 
\eeq
whence we may identify the Lagrange multiplier $\Omega$
with the angular velocity.  To understand the consequences
of the second condition
(vanishing of the $\delta\rho$ coefficient),
we need to interpret the $d\varepsilon/d\rho$ derivative.
If our $\delta$ variations are nondissipative, this
is an adiabatic derivative, and the first law of the thermodynamics
gives
\beq
{d\varepsilon\over d\rho} = {1\over\rho} ( \varepsilon + P) = H,
\eeq 
where $P$ is the gas pressure, and $H$ the enthalpy function.
The vanishing of the $\delta\rho$ coefficient then yields
\beq\label{fixE}
-{v^2\over2} + H + \Phi = {\zeta},
\eeq
whence we may identify the Lagrange multiplier ${\zeta}$
with the conserved Bernoulli constant. Equation (\ref{fixE})
is simply the integrated form of the hydrostatic equilibrium
equation for a uniformly rotating system
\beq
0= {1\over\rho} \del P + \del\left( \Phi -{R^2\Omega^2\over2}\right)
\eeq
(recall $dH = (1/\rho)dP$), 
showing that the equilibrium solution is an energy extremum.  We
do not know at this point that the extremum corresponds to a 
minimum.  For that, we must go to second order in the energy
variation.

\subsection{Second Order Variations}

The second-order variation in the integrand of $E$ is
\beq\label{secondord}
\delta^2E = \int 
{1\over 2} \left[ \rho (\delta v)^2 + 2 v \delta\rho\, \delta v
+ (\delta\rho)^2 {dH \over d\rho}\right]\, dV.
\eeq
Define the volume specific angular momentum $j = \rho R v$.
Then
\beq
\rho(\delta v)^2 + 2v\delta\rho\, \delta v =
{(\delta j)^2\over \rho R^2} - \rho v^2 \left(\delta\rho\over\rho\right)^2.
\eeq
Note additionally that 
\beq
{dH\over d\rho} = {1\over \rho}{dP\over d\rho} = {a^2\over \rho},
\eeq
where $a$ is the adiabatic sound speed, so that upon substituting
for the cross term $2 v \delta\rho\, \delta v$ in
(\ref{secondord}), we are brought to
\beq
\label{main}
\delta^2 E =
\int {\rho\over2}\left[
\left(\delta j\over \rho R\right)^2 + (a^2-v^2) \left(\delta\rho\over
\rho\right)^2\right]\, dV.
\eeq
Equation (\ref{main}) is the main result of this {\em Letter}.  
If the rotation is subsonic everywhere, then $\delta^2E > 0$,
and the original equilibrium is a true energy minimum.  But if the
rotation is supersonic at some point, it is possible to find
variations (say with $\delta j = 0$) for which $\delta^2 E < 0$.
In this case, the energy is not a local minimum, and the possibility
of instability is present.  Whether a given system is unstable or not
depends upon whether it is possible to reach the states of lower
energy.  Weak magnetic fields give fluids new internal degrees of
freedom, and may provide the pathway to these lower energy states.
Subsonic and supersonic rotation should have very
different consequences for the host system.   

\section {Discussion}

We are now in a position to answer the question posed in the {\em
Letter} title.  Subsonic uniform rotation is a true energy minimum,
whereas supersonic rotation carries no such guarantee.  Note that our
procedure does not ensure instability in the latter case, it simply
shows that neighboring states of lower energy exist.  If a weak
magnetic field is present, however, these states should be accessible.
Thus, we would expect the MRI to eliminate a small angular velocity
gradient and in subsonic rotation, but to accentuate it in supersonic
rotation.

The physical content of our result may be understood as follows.  The
most unstable disturbances (i.e., ``steepest descent'' in the $\delta
\rho\, \delta j$ plane) in a supersonically rotating gas cloud are
associated with $\delta j = 0$ disturbances.  While any particular
system may or may not be able to follow the route of steepest descent,
it is of interest to understand the reason that $\delta j = 0$
corresponds to such a path.  At a given radial distance from the axis,
the vanishing of $\delta j$ would mean that either (1) the density
rises as the angular velocity drops; or (2) that the density drops as
the angular velocity rises.  The first describes efficient extraction
of {\em mass specific} angular momentum and the development of a
pressure-supported slowly rotating core, the second describes the
development of an extended low density disk and the acquisition of mass
specific angular momentum.

Typical disk flow will of course rotate rapidly near the core and
slowly in the outer regions. But the $\delta j = 0$ states
infinitesimally close to the uniformly rotating equilibrium solution
should not be confused with some kind of quasi-Keplerian disk
profiles.  Rather, they represent the virtual state of the disk after
it is subject to variations but before it has relaxed to another
equilibrium.  However, this behavior is also recognizable as the
evolutionary strivings of real systems, with slowly rotating
pressure-supported dense cores and diffuse rotationally-supported
surroundings.

Self-gravity represents a significant mathematical complication to our
analysis, but not necessarily a physical one.  One expects the
bifurcation condition between supersonic and subsonic behavior to be
reasonably accurate even in self-gravitating systems, since it is the
outer regions that are prone to instability.  Here, the direct effects
of self-gravity tend to be locally less important.  A numerical
simulation is being undertaken by the authors to investigate this
point in detail.  

One practical application of these ideas is to be found in the scales
of rotating molecular cores internal to larger turbulent cloud
complexes.  Barranco \& Goodman (1998) and Goodman et al. (1998)
present the results of an observational study of coherence in molecular
cloud cores.  On length scales smaller than $R_{coh}\sim  0.1$ pc,
molecular linewidths are constant with scale, while on larger length
scales than $R_{coh}$ the linewidths are scale-dependent.  It is
suggested that $R = R_{coh}$ marks the transition between coherent and
noncoherent (i.e. turbulent) behavior, and is, as such, an inner scale
to a turbulent cascade.  The interesting question of what sets the
value of $R_{coh}$ remains open, with these authors favoring an
interpretation in which $R_{coh}$ is a dissipation scale for MHD
processes.

The results presented in this {\em Letter} may be relevant to
understanding the formation of these uniformly rotating cloud cores if
the existence of lower energy states for supersonic rotation follows
the scenario we have presented above.  The velocity gradient
$\Omega\sin i$ (where $i$ is the inclination of the rotation axis to
the line of sight) of several cloud cores is presented in Table (4) of
Barranco \& Goodman (1998).  A representative value for $\Omega$ from
this tabulation is about 1 km s$^{-1}$ pc$^{-1}$.  At $T=10$ K, the
adiabatic sound speed for molecular gas of cosmic abundances is 0.22 km
s$^{-1}$.  This means that the radius of the cores should all be less
than the sonic rotation radius $R_s= 0.22$ pc, and such is indeed the
case.  Moreover, since the cores have likely coalesced, their inferred
scale of 0.1 pc is quite consistent with this larger initial radius of
core detachment.  The above argument ignores magnetic and turbulent
support in the cores, but such support appears to be subsonic for the
$H_2$ gas in these regions (Barranco \& Goodman 1998).
In this interpretation, molecular cores do in fact mark the scale at
which a transition to coherence occurs, but this scale is present
for reasons that are essentially dynamical, not dissipative, in
character.
 
A somewhat more theoretical implication of our results concerns the
edge behavior of uniformly rotating cores.  A polytrope has a
temperature scaling as $T\sim \rho^{\gamma-1}$ ($\gamma$ is the usual
adiabatic index), so that any finite rotation speed for sufficiently
low density will become supersonic.  Edge cooling will have a similar
effect, more generally.  In essence, uniform rotation may not be
maintained across the radial extent of a gaseous body with a cold
outer surface layer.  Provided there is a pathway to the local
lower-lying energy states, the supersonic surface layers will soak up
angular momentum, and form an outer disk.  This process will continue
on a time scale given by the cooling time of the core's surface layers.
Keplerian-like disks with uniformly rotating cores would thereby
result.

A more definitive exploration of the evolutionary differences between
supersonic and subsonic rotation, including the effects of
self-gravity, is best carried out by way of numerical simulation
techniques.  These are currently in progress, and will be reported upon
in a forthcoming journal article.

\bigskip

We thank J.~Papaloizou and C.~Terquem for helpful discussions.  SF
would like to thank the Programme National de Physique Stellaire and
the Programme National de Plan\'etologie for travel support.  SAB is
grateful to the Institut d'Astrophysique de Paris for its visitor
support, and acknowledges support from NASA grants NAG 5--9266 and NAG
-- 10655, and NSF grant AST--0070979.

\end{document}